\documentclass[runningheads]{llncs}
\usepackage[T1]{fontenc}
\usepackage{array}
\usepackage{graphicx}%
\usepackage{multirow}%
\usepackage{amsmath,amssymb,amsfonts}%

\usepackage{mathrsfs}%
\usepackage[title]{appendix}%
\usepackage{xcolor}%
\usepackage{textcomp}%
\usepackage{manyfoot}%
\usepackage{booktabs}%
\usepackage{algorithm}%
\usepackage{algorithmicx}%
\usepackage{algpseudocode}%
\usepackage{listings}%

{\newcounter{normalfootc}
\renewcommand{\footnote}[1]{%
    \footlabel{footsaferefiwontuse\thenormalfootc}{#1}%
    \addtocounter{normalfootc}{1}%
}}
\newcommand{\amend}[1]{\noindent\textcolor{black}{#1}}

\begin{document}
\title{Research Knowledge Graphs: the Shifting Paradigm of Scholarly Information Representation}
\titlerunning{Research Knowledge Graphs for Scholarly Information Representation}
\author{Matth\"{a}us Zloch\inst{1,2}\ \and
Danilo Dess\`{i}\inst{3} \and
Jennifer D'Souza\inst{4} \and
Leyla Jael Castro\inst{5} \and
Benjamin Zapilko\inst{1} \and
Saurav Karmakar\inst{1} \and
Brigitte Mathiak\inst{1} \and
Markus Stocker\inst{4} \and
Wolfgang Otto\inst{1} \and
S\"{o}ren Auer\inst{4} \and
Stefan Dietze\inst{1,2}
}
\authorrunning{M. Zloch, D. Dess{\`{i}}, J. D'Souza, L. Castro, B. Zapilko, S. Karmakar et al.}
\institute{GESIS – Leibniz Institute for the Social Sciences, K\"oln, Germany \\
\email{firstname.lastname@gesis.org} \and
Heinrich-Heine-University, D\"usseldorf, Germany \\
\email{firstname.lastname@hhu.de} \and
Department of Computer Science, College of Computing and Informatics, University of Sharjah, UAE \\
\email{ddessi@sharjah.ac.ae} \and
TIB – Leibniz Information Centre for Science and Technology, Hannover, Germany \\
\email{firstname.lastname@tib.eu} \and
ZB MED – Information Centre for Life Sciences, K\"oln, Germany
\email{ljgarcia@zbmed.de}
}

\renewcommand*{\thefootnote}{\fnsymbol{footnote}}
\renewcommand*{\thefootnote}{\arabic{footnote}}

\maketitle
\begin{abstract}
Sharing and reusing research artifacts, such as datasets, publications, or methods is a fundamental part of scientific activity, where heterogeneity of resources and metadata and the common practice of capturing information in unstructured publications pose crucial challenges. Reproducibility of research and finding state-of-the-art methods or data have become increasingly challenging. In this context, the concept of Research Knowledge Graphs (RKGs) has emerged, aiming at providing an easy to use and machine-actionable representation of research artifacts and their relations. That is facilitated through the use of established principles for data representation, the consistent adoption of globally unique persistent identifiers and the reuse and linking of vocabularies and data. This paper provides the first conceptualisation of the RKG vision, a categorisation of in-use RKGs together with a description of RKG building blocks and principles. We also survey real-world RKG implementations differing with respect to scale, schema, data, used vocabulary, and reliability of the contained data. We also characterise different RKG construction methodologies and provide a forward-looking perspective on the diverse applications, opportunities, and challenges associated with the RKG vision.

\end{abstract}

\section{Introduction - Open Science Challenges}
\label{sec1}

The scientific process constantly produces and consumes scientific resources, such as publications, datasets, research methods, software, machine learning models, and discipline-specific research instruments. 
That has led to an ever-growing amount of research artifacts, where finding, understanding, and reusing them is crucial to scientists' daily activities. 
The widespread availability of diverse datasets and significant advancements in computational capabilities have led to the adoption of data science and artificial intelligence techniques across various research fields and disciplines.
Deep learning methods, in particular, have become the dominant approach in research areas such as natural language processing and image analysis. 
These methods typically involve a combination of code, machine learning models, and training data. 
The lack of transparency about dependencies and relations between such resources has led to a reproducibility crisis, where reproducing research and determining the state of the art has become increasingly challenging \cite{Dacrema2019}.
To help researchers find, share, and reuse such resources, there is a need for infrastructures  incorporating appropriate techniques to structure such information.
In recent years, a number of platforms and services have emerged, aiming at organizing different types of information. 
These include bibliographic databases such as DBLP\footnote{The dblp computer science bibliography,~\url{https://dblp.org}}, SCOPUS\footnote{Scopus - the abstract and citation database,~\url{https://www.scopus.com/}}, and Web of Science\footnote{\url{https://clarivate.com/products/scientific-and-academic-research/research-discovery-and-workflow-solutions/webofscience-platform/}}, 
dataset portals such as DataCite\footnote{\url{https://datacite.org/}}, Zenodo\footnote{\url{https://zenodo.org/}} or da$|$ra\footnote{\url{https://www.da-ra.de/}}, 
code sharing portals such as GitHub\footnote{\url{https://github.com}}, 
or bibliographic search engines such as Google Scholar\footnote{\url{https://scholar.google.com}}. 
While such services usually focus on artifacts of particular types (e.g., either publications or research datasets), \textit{dependencies} between artifacts of different types are crucial to better understand the \textit{provenance}, the \textit{context} of individual resources, and how they relate to each other, e.g., how a specific dataset was produced or what methods contributed to a specific finding. 
Such relations are usually not reflected as part of the aforementioned infrastructures.

Despite recent efforts by the research community to get datasets and software recognized as first-class citizens (e.g., cited on their own), unstructured scholarly publications remain the primary source of scientific insight. As a result, knowledge about scientific advancements and resource dependencies is often buried within these unstructured documents, with Portable Document Format (PDF) being the standard format for sharing such information. 
Extracting and utilizing this knowledge as part of the scientific process is a laborious and time-consuming task~\cite{otto-etal-2023-gsap}.

\begin{figure}
    \centering
    \includegraphics[width=\textwidth]{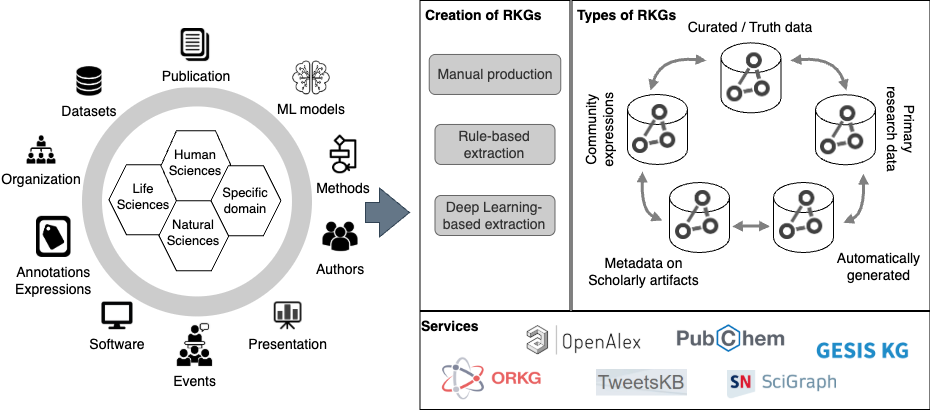}
    \caption{The figure illustrates examples of scholarly artifacts, methodologies to build RKGs, the five categories described in this paper, and examples of well-known services built on top.}
    \label{fig:figure1}
\end{figure}

To address these challenges, the research community has begun to design and develop Knowledge Graphs (KGs), i.e., networks of machine-readable, semantically rich, interlinked descriptions of entities and their relationships, usually expressed as graph-based databases either as Resource Descriptor Framework (RDF) triples or property graphs. To enhance the interoperability and understanding of Knowledge Graphs, researchers employ standardized vocabularies and ontologies, defining a common set of terms within a specific domain. This semantic layer facilitates consistent interpretation and communication. Furthermore, they use Persistent Identifiers (PIDs) providing a unique and persistent way to reference and retrieve specific resources and ensuring their long-term stability and accessibility. 

This paper investigates, describes, and categorizes KGs of research artifacts (e.g., methods, datasets, research papers) and entities (e.g., paper authors, organizations)~\cite{schindler2022_pubmed,dessi2022cs-kg,yaser2019ORKG} within the scholarly domain; we refer to them as Research Knowledge Graphs (RKGs). 
They stand as a transformative technology for the scholarly landscape, aiming in the long term at simplifying the way scientific outcomes are represented and utilized. 
Notable examples are OpenAlex~\cite{priem2022openalex}, the Microsoft Academic Graph~\cite{farber2019microsoft}, the GESIS Knowledge Graph\footnote{\url{https://data.gesis.org/gesiskg}}, and Springer SciGraph\footnote{\url{https://communities.springernature.com/users/82895-sn-scigraph}}, holding scholarly metadata information about research publications, research data, and authors. Another more recent example is the Open Research Knowledge Graph (ORKG)~\cite{yaser2019ORKG} that includes aspects such as research objectives and research problems. SoftwareKG \cite{schindler2020_softwarekg} is an RKG representing software usage and citations in scholarly works. Beyond these examples, RKGs like TweetsKB~\cite{fafalios_tweetskb_2018} and ClaimsKG~\cite{tchechmedjiev2019claimskg} offer a different focus, containing interlinked and semantically annotated research data. These RKGs serve research purposes, such as the long-term collection of tweets~\cite{fafalios_tweetskb_2018}, and the presentation of factual information alongside ratings extracted from fact-checking websites~\cite{tchechmedjiev2019claimskg,gangopadhyay2024investigating}. 

By representing relationships between research artifacts and scholarly domain entities through machine-actionable data structures~\cite{otto-etal-2023-gsap}, RKGs not only enhance data reusability but also empower machine-driven applications, encouraging new paths for scientific exploration and analysis. 
\amend{Well-structured RKGs enable researchers to trace data lineage, efficiently identify appropriate methods, and assess result validity. For publishers, RKGs offer structured representations of research contributions, facilitating indexing and discovery. Citizen scientists and open science initiatives similarly benefit from increased transparency and structured access to scholarly knowledge.}
One more key aspect is their compliance with the Findable, Accessible, Interoperable, Reproducible (FAIR) principles~\cite{wilkinson2016fair} which enhances the appeal of RKGs to diverse stakeholders interested in exploring, accessing, and making use of scholarly knowledge.
However, the development and usage of RKGs vary widely, originating from diverse data sources and employing different methodologies. This paper delves into this world, exploring RKGs' significance, applications, and challenges, to provide a comprehensive understanding of their pivotal role in advancing scholarly knowledge. Figure~\ref{fig:figure1} sketches the outlook our paper describes; on the left side the reader can observe scholarly entities and research artifacts RKGs target, on the right side, the existing types and nature, and in the middle the processes that can be used to generate them. The next sections further delve into these aspects. More precisely, to understand the variety of uses RKGs can offer  Section~\ref{sec:conceptualization-of-scholarly-knowledge} provides prominent examples of RKGs, analyzing their diverse nature. In Section~\ref{sec:methods-for-constructing-rkgs}, the focus shifts to the methodologies to build RKGs and highlights the related challenges. RKGs' role in reshaping the scholarly information representation paradigm is a topic delved further in Section~\ref{sec:challenges-and-perspectives}, which emphasizes the need for standardized approaches and outlooks future perspectives. Finally, Section~\ref{sec:conclusion} concludes the paper and remarks on the role of RKGs in the Scholarly Domain future.

\section{Conceptualization of Scholarly Knowledge}
\label{sec:conceptualization-of-scholarly-knowledge}
To provide an overview of the types of RKGs within the scholarly domain, the following categorization system will help us to characterize existing RKGs based on the dimensions of data quality, but also with regard to changes to the schema, data growth, vocabulary reuse, and graph inter-connectedness. \amend{The dimensions are grounded in observations of existing RKG implementations and their role in different scholarly infrastructural projects (institutions and products are mentioned in the corresponding sections).}
Table~\ref{tab:classification-long} shows an overview of the five categories and a comparison using RKGs' features over the five dimensions. 

\vspace{0.5em}
\noindent \textbf{1. Scholarly resource metadata}
Scholarly metadata resources and services serve as the primary access point for researchers to search for publications and associated artifacts. 
They facilitate bibliometric studies, perform research assessments, and effective research management. 
Managed by organizations focused on (i) providing access to preserved research artifacts, (ii) collecting, aligning, and aggregating metadata from various publishers, and (iii) investigating specific research aspects or use cases,
these resources are characterized by high-quality, curated content contributed by (a) researchers describing their work and (b) publishers ensuring proper quality control, indexing, and discoverability/findability of research artifacts. 
The schema used by these resources is stable, facilitating the development of research engines and services. 
However, the metadata vocabulary may vary depending on the organization and its focus. These resources expand over time and receive periodical updates. \amend{Validation is typically conducted by cross-referencing metadata with authoritative sources (e.g., CrossRef, ORCID) and leveraging expert curation.}
Examples of single-organization metadata resources include  \textit{SciGraph}\footnote{ \url{https://communities.springernature.com/users/82895-sn-scigraph}}, primarily encompassing SpringNature publications, and \textit{CultureGraph}\footnote{CultureGraph: \\\url{https://www.dnb.de/DE/Professionell/Standardisierung/AGV/\_content/culturegraph\_akk.html}}, provided by the \textit{Deutsche Nationalbibliothek}, linking library network metadata from Germany and Austria, along with the German National Library. 
Examples of aggregated resources are \textit{ResearchGraph}\footnote{ResearchGraph: \url{ https://researchgraph.org}}, a nonprofit metadata organization initiative closely aligned with the Research Data Alliance; \textit{OpenAIRE Research Graph}\footnote{OpenAIRE: \url{ https://graph.openaire.eu}} and OpenAlex\footnote{OpenAlex: \url{https://docs.openalex.org/}}, which provide integrated metadata about funders, organizations, researchers, research communities, and publishers; and \textit{PID Graph}\footnote{PID Graph: \url{https://api.datacite.org/graphql}}, which is a DataCite service that uses a GraphQL interface to enable integrated metadata searches on entities like datasets, publications, and people. 
Examples of metadata resources designed for more specific use cases are \textit{WikiCite/Scholia}\footnote{WikiCite/Scholia: \url{http://wikicite.org} and \url{https://scholia.toolforge.org/ }}, a Wikimedia project for organizing bibliographic information and researcher profiles for Wikipedia/Wikidata; and OpenCitations\footnote{OpenCitations: \url{https://opencitations.net/}}, which is an infrastructure dedicated to the publication of open bibliographic and citation data.
\amend{An example of a domain-specific RKG is the GESIS KG\footnote{\url{https://data.gesis.org/gesiskg/}} which comprises metadata of over 450,000 scientific resources (datasets, publications, variables, and survey instruments) from the social sciences and its semantic relationships in an integrated and consistent form and makes them accessible for reuse. This RKG serves also as a backbone for the GESIS Search\footnote{\url{https://search.gesis.org/}}.}

\vspace{1em}
\noindent \textbf{2. Quality-controlled ground truth data}
For many scholarly information extraction (IE) tasks, the availability of ground truth datasets is limited. Typically, publications related to scholarly IE tasks provide a ground truth dataset, a trained model on that dataset, and an automatically generated RKG. Alternatively, they might introduce a novel, fine-tuned model based on another pre-trained model, demonstrating superior performance through transfer learning.
Datasets in this category are crucial for quality assurance checks  during machine learning model training for downstream tasks. 
High-quality datasets in this category are usually annotated, curated \amend{, and validated} by humans (trained crowd-workers, students, postdocs, domain experts) and employ controlled metrics, such as Fleiss' $\kappa$ or Cohen's $\kappa$, to measure inter-annotator agreements. 
However, manual annotation is resource-intensive and time-consuming. Therefore, datasets in this category tend to be smaller and fewer compared to other categories. 
Their schemas remain fixed and stable, and the data remains static unless a new version is published under a (new) persistent identifier (PID) for reference purposes.
Whenever ground truth data is stored in an RKG (either natively or via programmatic generation), the data becomes more connected and meaningful thanks to the use of existing (controlled) vocabularies and PIDs that can later connect other RKGs. 
Notable examples of such ground truth data in the form of RKGs include Software Mentions in Science (SoMeSci)~\cite{schindler2021_somesci} and a specialized corpus for scientific literature entity tagging of tasks, datasets, and metrics, namely TDMSci~\cite{hou-etal-2021-tdmsci}.

\begin{table}[t!]
    \caption{Overview of the introduced qualitative categories for research knowledge graphs.}
    \centering
    \begin{tabular}{l|p{2.3cm}|p{1.5cm}|p{1.5cm}|p{1.4cm}|p{2.1cm}|p{2.4cm}}
        \hline
        & \textbf{Category} & \textbf{Scale} & \textbf{Schema} & \textbf{Data} & \textbf{Vocabulary} & \textbf{Connectedness} \\
        \hline
        1. & Scholarly resource metadata & medium & fixed and stable & evolving & well-defined, focused subset & high \\
        2. & Quality-controlled ground truths data & small & fixed and stable & stable & well-defined, focused subset & high \\
        3. & RKGs of primary research data & varies, potentially large & potentially fixed and stable & stable & focused subset & varies, potentially high \\
        4. & Community-expressions of scholarly artifacts & medium & evolving openly, template design & evolving & recommended/ used, not enforced & local, varies \\
        5. &  Automatically generated RKGs focusing on scholarly artifact relations & large (e.g., 300M) & fixed and stable & potentially evolving & focused subset & high \\
        \hline
    \end{tabular}
    \label{tab:classification-long}
\end{table}

\vspace{1em}
\noindent \textbf{3. RKGs representing and enriching primary research data}
\label{subsec:RKGs:primary-research-data}
Contrary to scholarly resource metadata, this category of research knowledge graphs contains primary research data (first-party data) collected directly by researchers, representing actual scientific domain knowledge.
These RKGs primarily serve researchers seeking specific research data pertinent to their discipline, such as sensor data from physical experiments or Twitter data for studying evolving opinions on topics in the Social Sciences domain.
Similar to automatically (NLP-)generated RKGs (last category), these RKGs undergo quality control to ensure transparency regarding the accuracy of specific information or features. Examples of validation methods include analysis of experimental results to ensure alignment with current research, expert review, and reproducibility checks.
The size of RKGs representing primary research data can be quite large, depending upon the domain and the scope of the data. 
Typically, the schema remains fixed and stable, with vocabulary reused whenever possible (e.g., NIF\footnote{\url{https://persistence.uni-leipzig.org/nlp2rdf/ontologies/nif-core/nif-core.html}}, DCAT\footnote{\url{https://www.w3.org/TR/vocab-dcat-2/}}, schema.org\footnote{\url{https://schema.org/}}). Access to these graphs is often provided via assigned and public DOIs.
The connectedness of such graphs (graph density) is influenced by the applied collection method and the utilization of the underlying vocabulary. 
Examples of research knowledge graphs containing primary research data can be found in ~\cite{fafalios_tweetskb_2018,LIU2023106679}.

\vspace{1em}
\noindent \textbf{4. Community-expressions of scholarly artifacts}
This category includes RKGs built with significant manual support from the scientific community, rather than primarily through automatic extraction from scholarly publications. However, those solutions offer automatic, NLP-powered services to aid users in uploading or entering semi-structured scientific content.
Their primary objective is to enhance the visibility and discoverability of scholarly knowledge, making them particularly valuable for researchers seeking (semi-automatically) access to scholarly information.
The nature of such graphs is cross-domain, facilitating the creation of tabular summaries (leaderboards) featuring state-of-the-art works across diverse disciplines. 
Unlike automated extraction methods requiring manual triggers for updates, community-driven RKGs tend to remain more up-to-date.
Nevertheless, such solutions may suffer from low quality in some sections due to community-driven vocabulary usage without quality assurance mechanisms. 
Continuous community involvement contributes to improving data quality over time. In fact, validation is obtained through consensus and crowd-sourced mechanisms such as post-publication peer review. 
Vocabulary (re-)use depends on additional services provided by the systems, resulting in a flexible and continually evolving schema, especially when new data is introduced. 
The level of entity inter-connectedness in such graphs can vary from low to high and may exhibit clustering.
An example of such a system is the previously mentioned ORKG~\cite{auer2020improving}, which aims to describe research papers in a structured way.

\vspace{1em}
\noindent \textbf{5. Automatically generated RKGs focusing on scholarly artifact relations}
\label{subsubsec:cat5-nlp-generated-scholarly-resource-relations}
This group comprises automatically generated RKGs, often produced through an NLP end-to-end pipeline, employing various deep learning technologies for tasks such as data cleaning, disambiguation, integration, named entity recognition and linking, and quality assessment. 
RKGs in this category focus on scholarly artifacts and their relationships. 
Typical use cases include employment in productive systems, such as scholarly search systems, for analyzing aspects like citation habits and patterns or competing solutions to NLP tasks. 
Users of these RKGs typically include researchers interested in scholarly information, or specific scholarly artifacts, like datasets, ML models, and software code repositories. 
Because of potential errors in automatic generation pipelines, the generated RKGs are in need of quality controls, in order to ensure transparency regarding the accuracy of specific information. %
As RKGs in this category can be created using pre-trained models fine-tuned on a variety of sources, they can scale to include hundreds of millions of triples. Validation is often performed using precision-recall analysis on top of smaller manually curated datasets and domain-expert verification of a portion of extracted relationships. 
The schema for these RKGs is typically fixed and stable, as schema elements are not necessarily dynamic in the pipeline processes.
Consequently, the vocabulary tends to concentrate on a particular subset of entity types like publications, software, models, and other research artifacts. While they may not be initially expressed in RDF, their structure and metadata can be easily annotated using vocabularies from schema.org, as well as formulated in accordance with schemas such as NIF and DCAT.
Notable examples of such research knowledge graphs include SoftwareKG~\cite{schindler2020_softwarekg,schindler2022_pubmed} and CS-KG~\cite{dessi2022cs-kg}.
\section{Methods for Constructing RKGs}
\label{sec:methods-for-constructing-rkgs}
Construction of RKGs relies on both manual effort and machine-based extraction and data lifting, where methods are applied and combined depending on the nature and characteristics of the RKG to be generated. While some entities, like articles and datasets, have readily available metadata that can be transformed to fit RKG requirements, extracting content from sources like Twitter or scientific literature is more complex. A particular focus of RKGs is on identifying and explicitly representing relations between different artifacts. Typical examples include dataset citations or method citations, that capture relations between datasets (methods) and publications or even between datasets and methods if citations are found in the same scholarly work~\cite{schindler2020_softwarekg,schindler2021_somesci,otto-etal-2023-gsap}. 

\vspace{0.8em}
\noindent \textbf{Manual RKG generation} For the case of scientific information originally expressed in scientific literature, manual production can occur pre- or post-publication of the original work while automated extraction occurs only post-publication. Post-publication manual production is ``crowdsource'' content, whereby contributors in the roles of authors or readers of original work manually produce structured expressions of the scientific information published originally in articles. By contrast, pre-publication manual production occurs by integrating the production of machine-actionable knowledge into the research data lifecycle, specifically data analysis, and can be heavily supported by infrastructure in order to automate most aspects of the task~\cite{boubakri2022orkg}.

\vspace{0.8em}
\noindent \textbf{Rule-based RKG extraction}
Generally, automated RKG construction involves turning unstructured scientific article text into structured entities and relations, and then republishing it in the form of graphs. The structuring can be based on an ontology or driven by NLP corpora. %
For instance, as an interdisciplinary RKG construction system, Research Spotlight~\cite{pertsas2018ontology} leverages the Scholarly Ontology (SO)~\cite{pertsas2017scholarly} which models research practices representing the information for ``who does what, when, and how'' in an RKG, for ontology-driven structured information extraction. The methodology is rule-based and applied on a database of scholarly articles. First, article metadata is mapped as SO instances via rules. The raw text undergoes sentence segmentation and entity extraction via an NER module. Relations are then generated using rules based on dependency trees, POS tags, SO semantic rules, and proximity constraints. Finally, URIs are created for the SO namespace and linked to relevant DBpedia entities for publication as linked data.

\vspace{0.8em}
\noindent \textbf{Deep Learning-based RKG extraction} Early systems followed a similar rule-based pipeline as discussed above, however with the advent of the ``deep learning tsunami''~\cite{manning2015computational} the KG construction pipelines incorporated neural learning methods. A generalized architecture is neural and symbolic involving a pipeline of complementary deep learning and rule-based solutions at various levels in the knowledge composition workflow. The CS-KG system~\cite{dessi2022scicero} is a relevant exemplar of neural-symbolic RKG construction. It integrates the DyGIEpp deep learning module~\cite{wadden2019entity}, which works within the transformers architecture~\cite{kenton2019bert} for predefined entity and relation extraction. Additionally, it uses the Computer Science Ontology classifier (CSO-C)~\cite{salatino2022cso} for further entity extraction. The Stanford Core NLP suite's OpenIE~\cite{angeli2015leveraging} determines open domain relations, and the Stanford POS Tagger~\cite{manning2014stanford} identifies verbs between entity pairs as relation candidates. As alternative or complementary components, the NER and relation extraction tasks for RKG construction can be addressed by finetuning transformer models based on the Bidirectional Encoder Representations from Transformers (BERT) architecture~\cite{kenton2019bert} on downstream task application corpora such as the NER or RE datasets discussed in Section 3. BERT offers pretrained parameters from large-scale general domain corpora such as Wikipedia or books trained with the masked language model objective producing language models capable of natural language understanding. The finetuning procedure then simply involves initializing the desired language model with the pretrained NLU models parameters and further in the context of task-specific architectures tuning the probabilistic parameters for a downstream extraction task given task-specific datasets. As such for the scientific domain, the widely used transformer language models are SciBERT~\cite{Beltagy2019SciBERT}, PubMedBERT~\cite{gu2021domain}, SemMedDB~\cite{kilicoglu2012semmeddb}, BioBERT~\cite{lee2020biobert}, BioClinicalBERT~\cite{alsentzer2019publicly}, and
BlueBERT~\cite{peng2019transfer}. In the realm of works around RKG construction, it is not uncommon to also leverage external knowledge bases as entity and relation extraction and linking candidates. This is the approach commonly witnessed in biomedicine. For instance, the iASiS knowledge graph~\cite{iasis_19} from biomedical data including scholarly publications on Lung Cancer and Dementia is generated utilizing NLP techniques coupled with the standardized biomedical ontologies such as UMLS~\cite{bodenreider2004unified} to annotate their extracted entities and relations. This latter system practically demonstrates the discovery of interactions between drugs in the treatments prescribed to lung cancer patients. As a final exemplar, in the domain of Biodiversity Science, OpenBioDiv~\cite{penev2019openbiodiv} is a knowledge graph from scholarly articles published by Pensoft\footnote{https://pensoft.net/} and Plazi\footnote{http://plazi.org/} that is structured by the OpenBiodiv-O ontology~\cite{senderov2018openbiodiv} for knowledge management. Thus, the themes of knowledge graphs vary between domain-specific subjects such as diseases in biomedicine or plant treatments in biodiversity, or domain-independent subjects such as research activity. While scientific articles are stored in silos isolated from each other, RKGs demonstrate how this is overcome by semantically combining different units of information.

\section{Outlook}
\label{sec:challenges-and-perspectives}

Representing data in the form of graphs can open new doors for better managing and making sense of produced research artifacts. This section looks at the RKGs' benefits and incentives and outlines their perspectives.

\vspace{0.8em}
\noindent \textbf{Research Management, FAIRness, and Consensus}
RKGs offer researchers and practitioners the possibility to describe research artifacts using established vocabularies and ontologies. These can enable several benefits, especially when well-established vocabularies are reused. For example, some RKGs model artifacts using schema.org, making them more visible, searchable, and findable to widely used search engines such as Google Search but also specific-purpose aggregators (e.g., the dataset search provided by the Australian Research Data Commons). In addition, well-established vocabularies also allow describing entities with rich metadata ensuring that this can be provided to the search engines for richer and better results. Standardized vocabularies simplify artifact update and maintenance while enabling clear and consistent research management (e.g., data and software management) and reducing the chances of introducing errors and inconsistencies. Finally, describing research artifacts with these technologies makes them more valuable due to a shared understanding and better integration and interoperability. Using well-established vocabularies with a broad scope, such as schema.org, already encouraged by recent RKGs e.g., SoMeSci and SoftwareKG, can promote effective governance by creating a common framework for data management, guaranteeing data quality, enabling findability and search on the web, and promoting community decision-making.

\vspace{0.5em}
\noindent \textbf{State-of-the-art Exploration and Enhanced Reproducibility}
Today, we are witnessing a reproducibility crisis and insufficient transparency is a growing concern: (i) many proposed solutions lack reproducibility due to unshared or unreported implementation details alongside traditional PDF publications, (ii) some machine learning models produce different outputs for the same input (e.g., generative large language models such as ChatGPT~\cite{openai2024chatgpt}) and cannot be explained due to their black-box nature, and (iii) keeping pace with the state-of-the-art is challenging; a large number of papers (and more and more also datasets and software) is published daily with no specific structure for searching specific relationships (e.g., find the performance of a given method on a specific dataset) leading several authors to adopt wrong baselines and claim themselves to be at the forefront of the field~\cite{Dacrema2019}. 
RKGs will be crucial to address these issues providing direct practical impact. 
First, RKGs can describe models along with their design parameters and results on certain tasks and datasets, facilitating their reproducibility (and to some extent transparency). 
Second, RKGs can provide explanations of black-box models, a direction already showing promising outcomes~\cite{balloccu2023reinforcement}. 
Finally, RKGs can bridge models with the tasks they are designed for, the datasets they are applied to, and the corresponding performance results, offering a comprehensive and connected global view of methods, tasks, data, and performance and easy access to the state of the art. For example, CS-KG can be used to i) discover uncommon scientific facts e.g., \texttt{<home automation}, \texttt{usesMaterial}, \texttt{telegram>}, ii) better understanding of computer science concepts e.g.,  \texttt{<ontology}, \texttt{supportsTask}, \texttt{reasoning>}, \texttt{<ontology}, \texttt{supportsTask}, \texttt{semantic interoperability>}, and iii) finding literature associated to a specific method and material such as \texttt{<graph neural network}, \texttt{solvesTask}, \texttt{molecular property prediction>}.
This will enable researchers to better compare their solutions as well as reviewers to provide a more solid basis for evaluations during the review process. By adopting RKGs, the research community can take significant strides toward research reliability, mitigating the reproducibility and state-of-the-art crisis. 

\vspace{0.5em}
\noindent \textbf{Research Knowledge Graph Integration}
RKGs are built with a focus on specific aspects or variables of interest and are suitable for specific use cases. As mentioned in the previous section, one advantage is their potential for interlinking, which widens RKG use cases and unlocks new knowledge generation. 
This feature sets RKGs apart from other data storage and usage technologies (e.g., databases or CSV files). 
RKGs, built using Semantic Web best practices, are web-accessible, utilize vocabularies and ontologies, employ non-proprietary formats, use permanent identifiers (PIDs) for entity identification, and enable links to internal and external research artifacts. 
Therefore, different RKGs can be explored using the same queries, simplifying information search across multiple RKGs and providing a common ground for interpreting results. 
Furthermore, several RKGs can represent the same entities but have different information about them. Linking RKGs allows users to collect various knowledge pieces about the same entity. For example, a user can first discover an author's papers from an RKG describing paper metadata, find links to NLP or community-expression RKGs describing the paper content, and discover the methods or dataset the author used. Interlinking RKGs opens the door to discovering a vast amount of new knowledge that is currently difficult to achieve due to the article-centric publishing paradigm.

\vspace{1em}
\noindent \textbf{Citability and Creditability of Research Artifacts }
Today, scholarly papers remain the most common form of referenced research artifacts ensuring their permanence in literature. There is an increasing demand to reference and cite other essential research artifacts. For example, ORCIDs are identifiers associated with researchers, DOIs (and PIDs) are used to identify datasets and software, and Research Organization Registry\footnote{\url{https://ror.org/}} (ROR) are used to identify organizations. However, many other research artifacts and entities require proper reference to enable their citability and credit to their owners. These encompass a wide range, such as machine learning models, methodologies, achieved results, and more. In this context, RKGs and their best practices in PIDs, i.e., long-lasting stable URIs associated with entities, can serve for example as a tool for reliably and efficiently attributing researchers, such as identifying them as owners, developers, or maintainers of a particular machine learning model. Furthermore, researchers can be acknowledged for their role in methodologies, and the methodologies themselves can be cited. Moreover, PIDs enable the linking of methodologies to tasks they solve, thus allowing the generation of valuable connections among different research artifacts. RKGs can make research interconnected and traceable, thus paving the way for a more liable and integrated scholarly research paradigm. 

\vspace{1em}
\noindent \textbf{Research Knowledge Graphs to Support Large Language Models}
With the recent advent of large language models (LLMs), researchers and professionals in various industries are increasingly incorporating this technology into their daily work routines. 
However, despite their impressive performance at first glance, LLMs come with several well-known limitations. 
These limitations include their high training and operational costs, challenges associated with maintenance and updates, tendencies to generate hallucinatory and inconsistent responses (depending on the language employed), and difficulties in tracing and attributing the sources of their answers. This is also problematic for the scholarly domain where LLMs can suggest fictional articles as well as misleading statistics (e.g., an erroneous citation count)~\cite{meloni2023integrating}. 
Furthermore, LLMs are static in nature, as they are trained only once at a specific point in time, which hinders their ability to continuously evolve and adapt to the changing knowledge, particularly in scholarly domains where research is in constant flux, with researchers continuously adding, modifying, or removing findings.

In this context, RKGs take a pivotal role and serve as a verifiable source of truth to obtain up-to-date information. 
RKGs significantly enhance the capabilities of LLMs by facilitating the integration of language-independent knowledge, enabling fact verification, enhancing contextual understanding, providing domain-specific expertise, enabling personalization, and adeptly addressing complex queries.

\section{Conclusion}\label{sec:conclusion}
This paper introduced RKGs as a transformative paradigm for storing, managing, and sharing research artifacts. By exploring types and construction methodologies underlying RKGs, we have outlined the transformative potential RKGs hold for scholarly information representation. 
Their adoption implies a profound shift in how scientists can conceptualize, organize, and interact with scholarly knowledge. 
By integrating disparate data sources and forming meaningful relationships, RKGs foster interdisciplinary collaboration and knowledge discovery. 
Scholars, researchers, and practitioners can benefit from linked data to explore complex research questions, identify emerging trends, and gain deeper insights into academic knowledge. 
Embracing this novel paradigm helps navigate the expanding knowledge landscape. 
Continued research and development of RKGs will refine our understanding and utilization of research artifacts unlocking new opportunities to address scientific challenges.

\subsubsection*{Acknowledgements.} This work is partially supported by the DFG-funded project NFDI for Data Science \& Artificial Intelligence (NFDI4DS, project number 460234259).

\bibliographystyle{splncs04}
\bibliography{llncs}

\end{document}